# UBIQUITOUS SMART HOME SYSTEM USING ANDROID APPLICATION

Shiu Kumar

Department of Information Electronics Engineering, Mokpo National University, 534-729, Mokpo, South Korea

## ABSTRACT

*This paper presents a flexible standalone, low cost smart home system, which is based on the Android app communicating with the micro-web server providing more than the switching functionalities. The Arduino Ethernet is used to eliminate the use of a personal computer (PC) keeping the cost of the overall system to a minimum while voice activation is incorporated for switching functionalities. Devices such as light switches, power plugs, temperature sensors, humidity sensors, current sensors, intrusion detection sensors, smoke/gas sensors and sirens have been integrated in the system to demonstrate the feasibility and effectiveness of the proposed smart home system. The smart home app is tested and it is able successfully perform the smart home operations such as switching functionalities, automatic environmental control and intrusion detection, in the later case where an email is generated and the siren goes on.*

## KEYWORDS

## 1. INTRODUCTION

With the continuous growth of mobile devices in its popularity and functionality the demand for advanced ubiquitous mobile applications in people's daily lives is continuously increasing. Utilizing web services is the most open and interoperable way of providing remote service access or enabling applications to communicate with each other. An attractive market for home automation and networking is represented by busy families and individuals with physical limitations.

IoTs can be described as connecting everyday objects like smart phones, internet televisions, sensors and actuators to the internet where the devices are intelligently linked together to enable new forms of communication amongst people and themselves [1]. The significant advancement of IoTs over the last couple of years has created a new dimension to the world of information and communication technologies. The advancement is leading to anyone, anytime, anywhere (AAA) connectivity for things with the expectation being that this extend and create an entirely advanced dynamic network of IoTs. The IoTs technology can be used for creating new concepts and wide development space for smart homes in order to provide intelligence, comfort and improved quality of life.

Smart home is a very promising area, which has various benefits such as providing increased comfort, greater safety and security, a more rational use of energy and other resources thus contributing to a significant savings. This research application domain is very important and will increase in future as it also offers powerful means for helping and supporting special needs of the elderly and people with disabilities [2], for monitoring the environment [3] and for control. There are a number of factors that needs to be considered when designing a smart home system. The system should be affordable, scalable so that new devices can be easily integrated into the system, and it should be user friendly [4].

With the dramatic increase in smart phone users, smart phones have gradually turned into an all-purpose portable device and provided people for their daily use. In this paper, a low cost wireless controlled smart home system for controlling and monitoring the home environment is presented. An embedded micro-web server with real IP connectivity is used for accessing and controlling appliances and other devices remotely from an Android based app, which can be used from any Android supported device. The Arduino Ethernet is used for the micro web-server thus eliminating the use of PC and the system requires user authentication in order to access the smart home system. Voice activation for switching applications has also been incorporated to aid users especially for the elderly and the disabled persons.

The remainder of the paper is organized as follows. In Section 2, a brief discussion of the related work is provided. The overall system architecture, implementation and the features of the proposed smart home system are presented in Section 3. Finally the conclusion with some further prospective works is presented.

## 2. RELATED LITERATURE

Smart home is not a new term for science society however, it is still far more away from people's vision and audition. As electronic technologies are converging, the field of home automation is expanding. Various smart systems have been proposed where the control is via Bluetooth[5-10], internet [11-13], short message service (SMS) based [14], etc. Bluetooth capabilities are good and most of current laptop/notebook, tablets and cell phones have built-in adaptor that will indirectly reduce the cost of the system. However it limits the control to within the Bluetooth range of the environment while most other systems are not too feasible to be implemented as low cost solution.

In [15], Wi-Fi based home automation system is presented. It uses a PC (with built in Wi-Fi card) based web server that manages the connected home devices. The users can manage and control the system locally (LAN) or remotely (internet). The system supports a wide range of home automation devices like power management components and security components. A similar architecture is proposed in [16] where the actions are coordinated by the home agent running on a PC. Other papers such as [17-20] also presented internet controlled systems consisting of a dedicated web server, database and a web page for interconnecting and managing the devices. These systems utilize a PC which leads to a direct increase in cost and power consumption. On the other hand, the development and hosting of the web page will also result in additional costs.

The design and implementation of a microcontroller based voice activated wireless automation system is presented in [21]. The user speaks the voice commands through a microphone, which is processed and sent wirelessly via radio frequency (RF) link to the main control receiver unit. Voice recognition module is used to extract the features of the voice command. This extracted signal is than processed by the microcontroller to perform the desired action. The drawback is that the system can only be controlled from within the RF range. Reference [22] also presents a voice activated smart home automation system. This system provides graphical user interface (GUI) using Microsoft Visual Basic software hosted by a PC, and uses Microsoft Speech Recognition engine. The signal is than transmitted via RF link to the microcontroller to which the home appliances are interfaced. Again a PC is used that account for an increased cost and power consumption.

International Journal of Computer Networks & Communications (IJCNC) Vol.6, No.1, January 2014

A significant contribution to smart home system has been made by the above mentioned systems. However, a PC is used as a server that increases the cost and power consumption while others require web page hosting that adds up the extra cost. The voice activation systems either use PC software or separate voice recognition module for speech recognition.

## 3. SYSTEM DESIGN

### 3.1. System Architecture

In the proposed design, a low cost smart home system for remotely controlling and monitoring the smart home environment is presented. An overview of the proposed system architecture is shown in Figure 1. The system consists of an app developed using the Android platform and an Arduino Ethernet based micro web-server. The Arduino microcontroller is the main controller that hosts the micro web-server and performs the necessary actions that needs to be carried out. The sensors and actuators/relays are directly interfaced to the main controller. The smart home environment can be controlled and monitored from a remote location using the smart home app, which will communicate with the micro web-server via the internet. Any internet connection via Wi-Fi or 3G/4G network can be used on the user device.

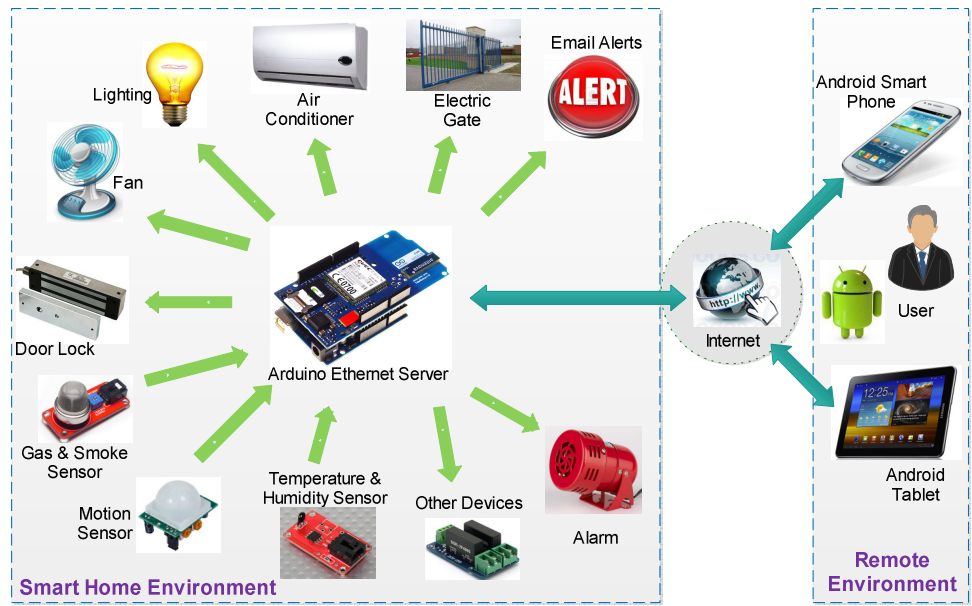

Figure 1. System Architecture of the proposed Ubiquitous Smart Home

The features that the proposed design offers are the control of energy management systems such as lightings, power plugs and HVAC (heating, ventilation and air conditioning) systems; security and surveillance system such as fire detection and intrusion detection with siren and email notifications; automatic smart home environment control such as maintaining a certain room temperature; voice activation for switching functions and has user authentication to access the smart home system.

### 3.2. Software development of the Android platform app

There are several platforms for developing smart phone applications such as Windows Mobile, Symbian, iOS and Android. In the proposed system, the Android platform app is developed as most of the phones and handy devices support Android OS. Java programming language using the Android Software Development Kit (SDK) has been used for the development and implementation of the smart home app. The SDK includes a complete set of development tools such as debugger, libraries, a handset emulator with documentation, sample code and tutorials. Eclipse (running on Windows 7 development platform), which is the officially supported integrated development environment (IDE) has been used on in conjunction with the Android Development Tools (ADT) Plug-in to develop the smart home app. The screenshots of the smart home app developed is shown in Figure 2 while the processing of the smart home app is shown in Figure 3.

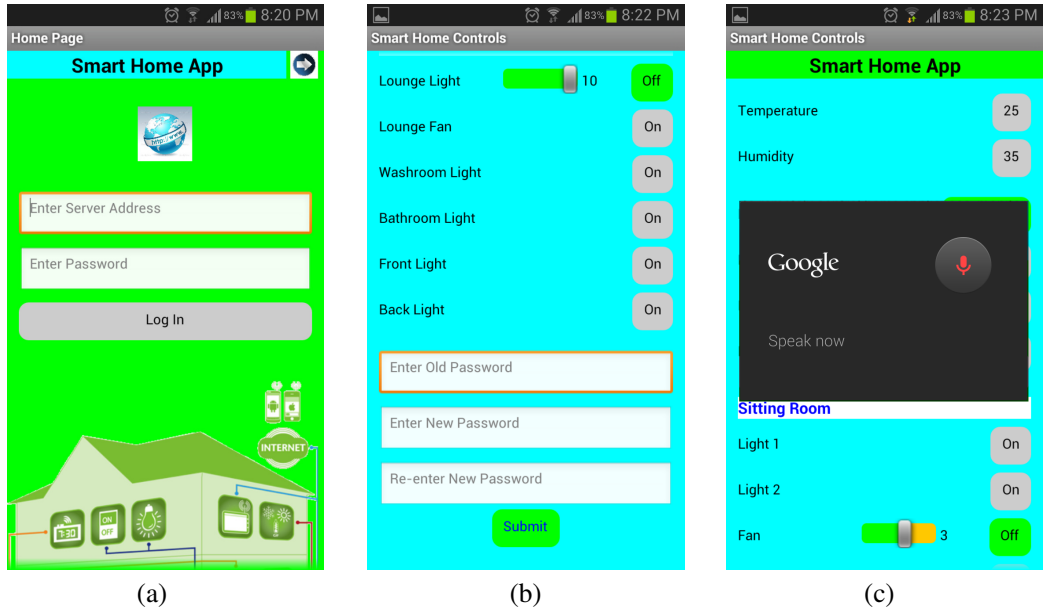

Figure 2. Screenshots of the proposed smart home app

The designed app for the smart home system provides the following functionalities to the user:

- Remote connection (via internet) to the smart home micro web-server; require server real IP and user authentication.
- Device control and monitoring.
- Scheduling tasks and setting automatic control of the smart home environment.
- Password change option.
- Supports voice activation for switching functions.

In order to successfully connect and access the smart home micro web-server, the user has to enter the correct real IP address and password (see Figure 2.a). If the micro web-server grants access to the smart home app, response packet containing response code 200 will be received. The app processes the response packet to determine the micro web-server's response. Response code 200 indicates the password is correct, and the app will switch to the main control page and synchronize using the data from the response packet to reflect the real time statuses of the smart home devices (see Figure 2.b). If the password is incorrect, response code 404 will be received. The general response packet layout is shown in Figure 4. The response code and devices with their statuses are separated by a space while the device and its status is separated by a colon (:). For example when the action requested by the user from the app to turn on Light 1 is successful, the response packet will be "200 Light_1:1". A zero indicates off state while a one indicates on state for the status for switching functions.

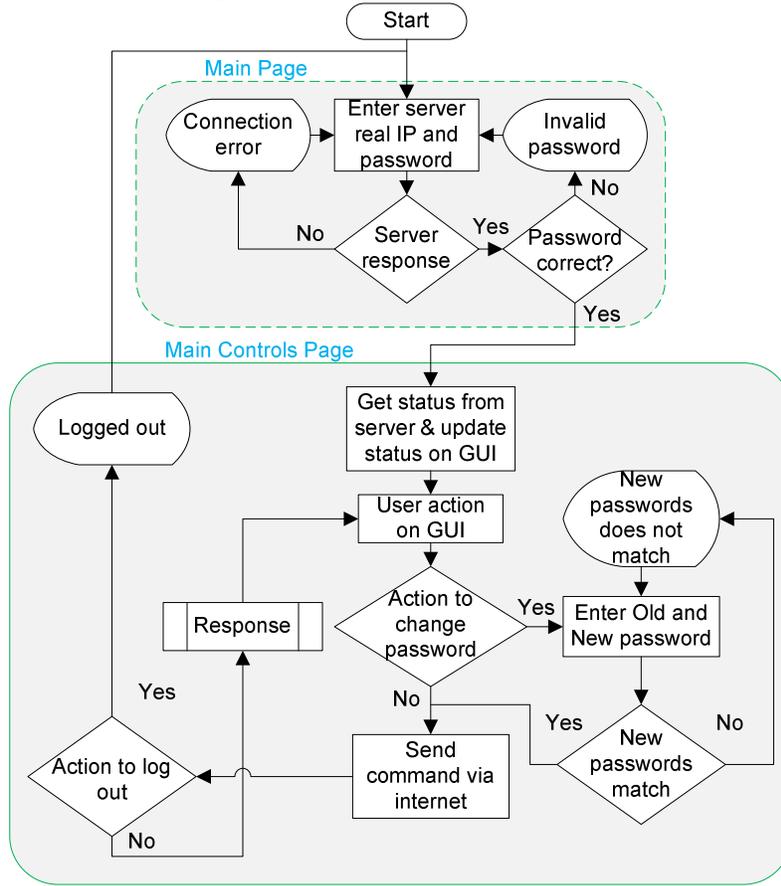

Figure 3. General processing of proposed smart home app

Figure 4. General packet layout of the response packet.

The user can perform the desired action from the GUI one's access is granted. Switching actions can also be performed through voice activation that uses the Google Speech Recognition engine available on the device (see Figure 2.c). The password can also be managed by the user from the smart home app. Clicking the password change button on the GUI will then require the user to enter the old and password (see Figure 2.b). If the new passwords match than the command packet containing the new password is sent to the micro web-server. If password is successfully changed, response code 201 will be received. Automatic mode can also be activated where the smart home environment will be controlled automatically, for example maintaining a certain room temperature and turning on/off certain light during night/day.

When the user performs an action on the smart home app, command packet is sent to the micro web-server via the internet. The general layout of the command packet is shown in Figure 5. The command packet if formatted in such a way that micro web-server is easily able to read and extract the information from the packet. For example for turning on the fan with the default

International Journal of Computer Networks & Communications (IJCNC) Vol.6, No.1, January 2014

password, the command packet sent will be "$1234$Fan_On" and for setting the fan speed to 2 the command packet will be "$1234$FanSpeed_2".

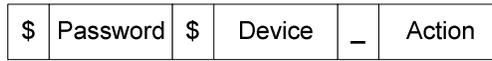

Figure 5. General packet layout of the command packet.

### 3.3. Software development of the smart home micro web-server

The main controller hosting the micro web-server acts as the heart of the smart home system consisting of the server application software and the Arduino microcontroller firmware. The server application software is the library implementation of the micro web-server running on the Arduino Mega 2560 using the Arduino Ethernet shield connected to the internet over TCP/IP, which can act as both the server and client. The Ethernet library "<Ethernet.h>" is used to send and receive data in conjunction with the microcontroller. The output messages sent to the smart home app is in JavaScript Object Notation (JSON) format.

Utilizing Web services is the most open and interoperable way of providing access to remote services or for enabling applications to communicate amongst each other. Simple Object Access Protocol (SOAP) and Representative State Transfer (REST) are the two classes of Web services. However, REST ful based Web service has been employed due to its light-weight compared to the SOAP based Web service offering similar functionalities. Standard GET and POST request operations have been utilized for communication between the smart home app and the micro web-server. For example, if the user action on the smart home app is to activate the house alarm system then the message exchange taking place is illustrated in Figure 6.

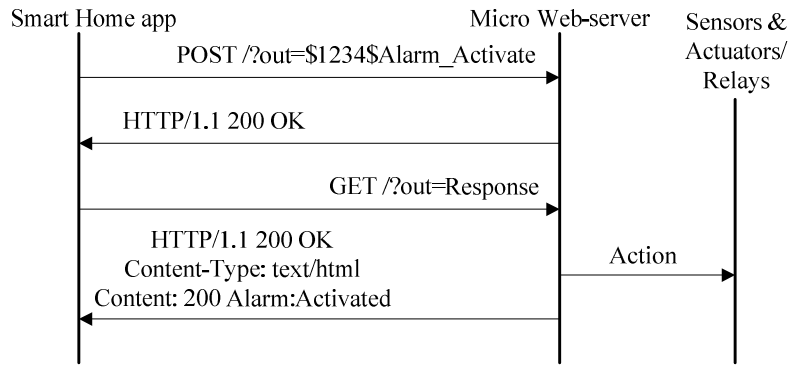

Figure 6. Message exchange between the smart home app and micro web-server

The main controller starts off by initializing the Ethernet in order to establish connection between them. Variables that also include the status variable are initialized and the interrupt on the microcontroller is enabled. If the main controller is run for the first time with the smart home server codes, then the default password of 1234 is initialized. Otherwise, the password is read from the EEPROM and stored in the password variable. The password is stored in the microcontrollers EEPROM so that it is still available even if the main controller starts after a complete shutdown, which can be due to reasons such as low or no power supply or the user restarts the system. Upon initializing, the system enters idle mode until a client is made available. Figure 7 shows the general flowchart for the main controller. When clients are made available, the command packet is received and the information is extracted.

Once the command packet is received and the data is extracted or decoded then the perform operation sub-routine that handles the authentication check and action to be carried out. Figure 8 shows the flowchart for the performance operation sub-routine.

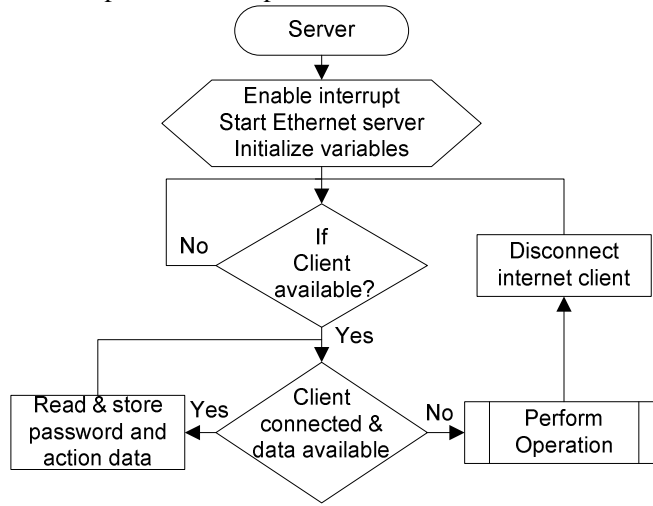

Figure 7. The flowchart showing the main controller operation

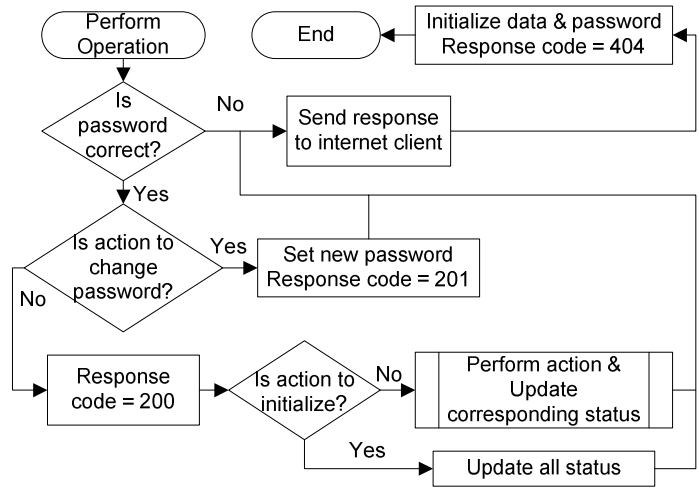

Figure 8. The main controller's Perform Operation sub-routine

Once the perform operation sub-routine is called, the user authentication that is the password is verified. If the password is incorrect, the response is sent back to the client/user app and the client is disconnected. However, if the authentication is correct then the action requested is performed. The response codes sent for different operations performed has already been discussed. If action requested is to change the password then new password is written to the EEPROM and the password variable is initialized to this new password. After every action is performed, the response packet is sent to the client after which, the response code is initialized to 404.

The microcontroller's interrupt is used to overlook security and surveillance such as smoke, fire, gas or intrusion detection. If any of the sensors are triggered, interrupt is generated. The microcontroller then determines the type of interrupt that have been generated. Once the type of interrupt generated is known, the siren is turned on and email notifications are sent to the user specified personals including the user. For example, if a fire is detected in the kitchen the email

International Journal of Computer Networks & Communications (IJCNC) Vol.6, No.1, January 2014

received by the user will have email subject as "Smart Home Alert" and the body as "Fire Detected in the Kitchen". The user can switch off the siren from the user app or it will turn off automatically after a certain period of time depending on user requirements.

### 3.4. Proposed smart home devices

The Arduino Mega 2560 and the Arduino Ethernet shield have been used to implement the smart home micro web-server. Arduino is an open-source electronics prototyping platform based on flexible, easy-to-use hardware and software. The Arduino Mega 2560 microcontroller board is based on the ATmega256 having 54 digital input/output pins. The Ethernet is interfaced to the Arduino via the Arduino SPI pins. Low voltage switching relays were used to integrate the devices with the Arduino for demonstrating the switching functionality. The LM35 temperature sensor was used to monitor the smart home environment. The CG313 MQ5 Smoke Gas Sensor Module for Arduino has been used for smoke detection. For intrusion detection, the HC-SR501 Infrared PIR Motion Sensor Module for Arduino has been utilized.

## 4. RESULTS AND DISCUSSION

The smart home system proposed in this paper was fully developed and tested to demonstrate its feasibility and effectiveness. The screenshots of the smart home app developed has been presented in Figure 2, which is described in Section 3.2. As mentioned, authentication is required to access the smart home system. Figure 9.a shows the message displaying that an invalid password was entered. This message is based on the response received from the smart home micro web-server. If correct authentication is provided, the app then proceeds to display the smart home controls page with a message notifying success of login as shown in Figure 9.b. When the voice activation function is used, if no command is captured, a message is displayed prompting the user to speak again (see Figure 9.c).

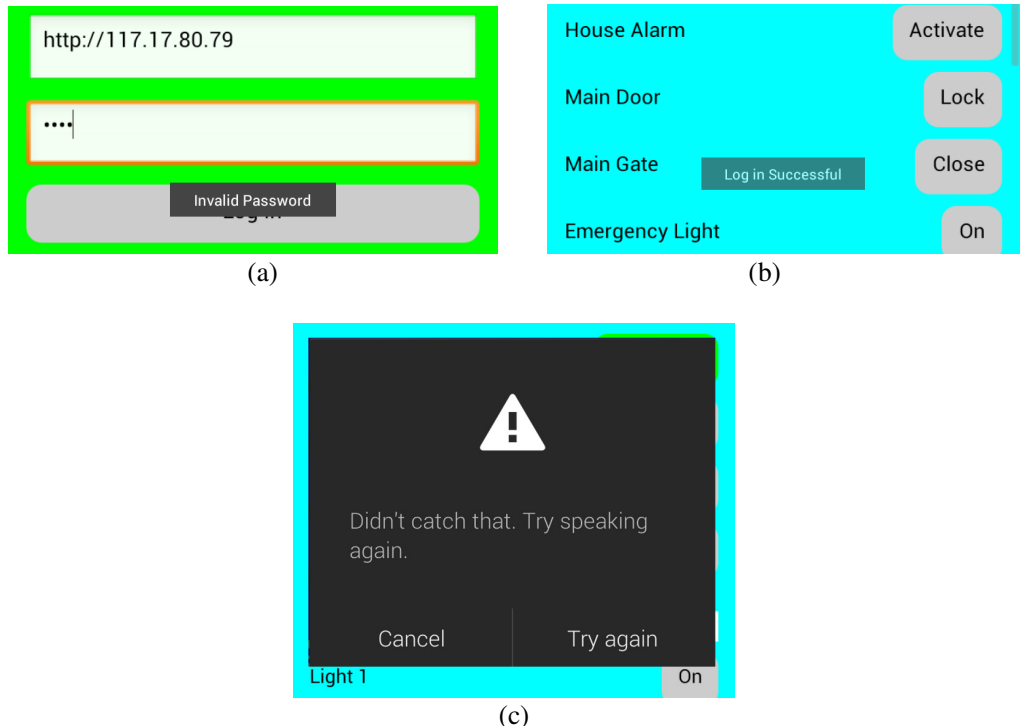

Figure 9. Screenshots of the proposed smart home app: (a) when invalid password is entered (b) when login is successful (c) when no voice command is captured.

The smart home system was fully functional for the switching applications and as the appliances are switched on the user interface is updated to reflect the current status. The smart home system was also tested for intrusion and fire detection whereby it successfully detected the respective events generating an email to the user and turning on the siren. Figure 10 shows the email generated (due to a fire detected in the kitchen) and received by the user on the mobile device and on the desktop PC where email was configured on Microsoft outlook.

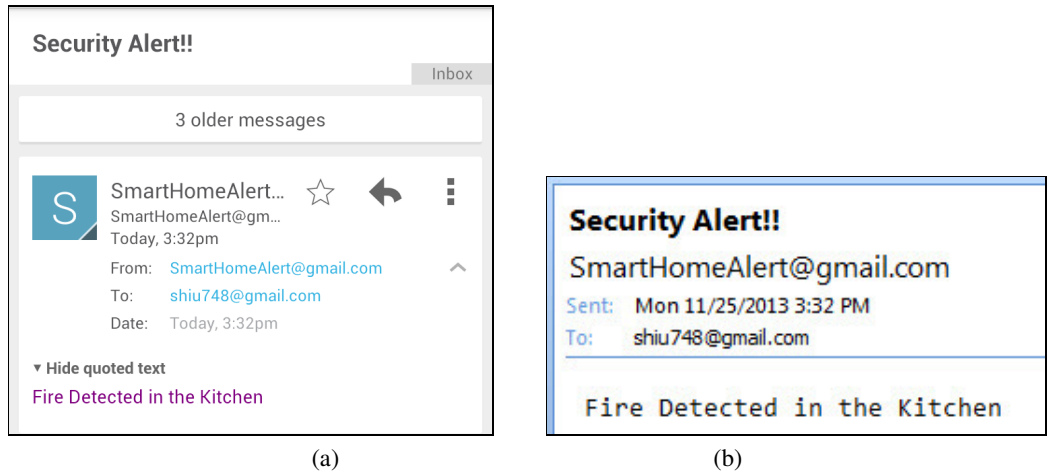

(a)                  (b)

Figure 10. Alert notification generated via email received: (a) on the android mobile device and (b) on the desktop PC

The proposed system has all the feature with respect to the research's mentioned in Section 2. On the other hand, it also has security features such as user authentication for accessing the smart home system, and intrusion and fire detection with alert notification. The system does not require a dedicated PC or voice processing module for the voice activation feature making it low cost and affordable.

## 5. CONCLUSIONS

In this paper, an internet based smart home system that can be controlled remotely upon user authentication is proposed and implemented. The Android based smart home app communicates with the micro web-server via internet using the REST ful based web service. Any android supported device can be used to install the smart home app, and control and monitor the smart home environment. A low cost smart home system has been developed which does not require a PC as all processing is handled by the microcontroller. The system also uses the Google speech recognition engine thus eliminating the need for an external voice recognition module. Prospective future works include incorporating SMS and call alerts, and reducing the wiring changes for installing the proposed system in pre-existing houses by creating a wireless network within the home environment for controlling and monitoring the smart home environment.

## ACKNOWLEDGEMENTS

Financial support for this study was provided by a grant from the Woojung Education and Culture Foundation. The author wishes to thank the Woojung Education and Culture Foundation for the kind and generous support, which helped in the purchasing of the research materials for conducting this research.

## REFERENCES

[1]   G. Kortuem, F. Kawsar, D. Fitton, and V. Sundramoorthy, "Smart objects as building blocks for the internet of things," *Internet Computing, IEEE,* vol. 14, pp. 44-51, 2010.

[2]   R. J. C. Nunes and J. C. M. Delgado, "An Internet application for home automation," in *10th Mediterranean Electrotechnical Conference (MELECON 2000)*, Lemesos, 2000, pp. 298-301.

[3]   F. Kausar, E. A. Eisa, and I. Bakhsh, "Intelligent Home Monitoring Using RSSI in Wireless Sensor Networks," *International Journal of Computer Networks & Communications,* vol. 4, pp. 33-46, 2012.

[4]   R. Piyare and M. Tazil, "Bluetooth Based Home Automation System Using Cell phone," in *IEEE 15th International Symposium on Consumer Electronics*, Singapore 2011, pp. 192 - 195.

[5]   S. Anwaarullah and S. V. Altaf, "RTOS based Home Automation System using Android," *International Journal of Advanced Trends in Computer Science and Engineering,* vol. 2, pp. 480-484, January 2013 2013.

[6]   C. Chiu-Chiao, H. C. Yuan, W. Shiau-Chin, and L. Cheng-Min, "Bluetooth-Based Android Interactive Applications for Smart Living," in *2nd International Conferenceon Innovations in Bio-inspired Computing and Appplications (IBICA 2011)*, 2011, pp. 309-312.

[7]   D. Javale, M. Mohsin, S. Nandanwar, and M. Shingate, "Home Automation and Security System Using Android ADK," *International Journal of Electronics Communication and Computer Technology (IJECCT),* vol. 3, pp. 382-385, March 2013 2013.

[8]   J. Potts and S. Sukittanon, "Exploiting Bluetooth on Android mobile devices for home security applications," in *Southeastcon, 2012 Proceedings of IEEE* Orlando, FL 2012.

[9]   R. A. Ramlee, M. H. Leong, R. S. S. Singh, M. M. Ismail, M. A. Othman, H. A. Sulaiman*, et al.*, "Bluetooth Remote Home Automation System Using Android Application," *The International Journal of Engineering And Science,* vol. 2, pp. 149-153, 11, January 2013 2013.

[10]  M. Yan and H. Shi, "Smart Living Using Bluetooth Based Android Smartphone," *International Journal of Wireless & Mobile Networks,* vol. 5, pp. 65-72, February 2013 2013.

[11]  C. C. Ko, B. M. Chen, S. Hu, V. Ramakrishnan, C. D. Cheng, Y. Zhuang*, et al.*, "A web-based virtual laboratory on a frequency modulation experiment " *IEEE Transactions on Systems, Man, and Cybernetics, Part C: Applications and Reviews,* vol. 31, pp. 295-303, 2001.

[12]  N. Swamy, O. Kuljaca, and F. L. Lewis, "Internet-based educational control systems lab using NetMeeting " *IEEE Transactions on Education,* vol. 45, pp. 145-151, 07 August 2002 2002.

[13]  K. K. Tan, T. H. Lee, and C. Y. Soh, "Internet-based monitoring of distributed control systems - An undergraduate experiment," *IEEE Transactions on Education,* vol. 45, pp. 128-134, May 2002 2002.

[14]  M. S. H. Khiyal, A. Khan, and E. Shehzadi, "SMS Based Wireless Home Appliance Control System (HACS) for Automating Appliances and Security," *Issues in Informing Science and Information Technology,* vol. 6, pp. 887-894, 2009.

[15]  A. ElShafee and K. A. Hamed, "Design and Implementation of a WiFi Based Home Automation System," *World Academy of Science, Engineering and Technology,* vol. 68, pp. 2177-2180, 2012.

[16]     R. D. Caytiles and B. Park, "Mobile IP-Based Architecture for Smart Homes," *International Journal of Smart Home,* vol. 6, pp. 29-36, 2012.

[17]     A. Z. AAlkar and U. Buhur, "An internet based wireless home automation system for multifunctional devices," *IEEE Transactions on Consumer Electronics,* vol. 51, pp. 1169-1174, 2005.

[18]     N.-S. Liang, L.-C. Fu, and C.-L. Wu, "An integrated, flexible, and Internet-based control architecture for home automation system in the Internet era," in *IEEE International Conference on Robotics and Automation*, Washington, DC 2002, pp. 1101 - 1106

[19]     A. Rajabzadeh, A. R. Manashty, and Z. F. Jahromi, "A Mobile Application for Smart House Remote Control System," *World Academy of Science, Engineering and Technology,* vol. 62, 2010.

[20]     U. Sharma and S. R. N. Reddy, "Design of Home/Office Automation Using Wireless Senosr Network," *International Journal of Computer Applications,* vol. 43, pp. 53-60, 2012.

[21]     K. P. Dutta, P. Rai, and V. Shekher, "Microcontroller Based Voice Activated Wireless Automation System," *VSRD Internation Journal of Electrocal, Electronics & Communication Engineering,* vol. 2, pp. 642-649, 2012.

[22]     M. R. Kamarudin, M. A. F., and M. Yusof, "Low Cost Smart Home Automation via Microsoft Speech Recognition," *International Journal of Engineering & Computer Science,* vol. 13, pp. 6-11, June 2013.

**Authors**

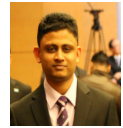

**Shiu Kumar** received Bachelor of Engineering Technology and Postgraduate Diploma in Electrical and Electronics Engineering from the University of the South Pacific in year 2009 and 2012 respectively. Currently he is pursuing Masters Degree in Electronics at Mokpo National University, South Korea. His research interests include Automation and Control, Wireless Sensor Networks, Embedded Microprocessor Applications, Artificial Intelligence and  Digital Electronics.